\documentstyle[11pt,newpasp,twoside,epsf]{article}
\def\edcomment#1{\iffalse\marginpar{\raggedright\sl#1\/}\else\relax\fi}
\reversemarginpar

\begin{document}
\title{Polarisation Observations at Jodrell Bank Observatory}
 \author{D. Athanasiadis$^{1}$, M. Kramer$^{1}$, I. H. Stairs,$^{2}$,
 A. G. Lyne$^{1}$} 
\affil{$^{1}$Jodrell Bank Observatory, Dept. of Physics \& Astronomy, The
 University of Manchester, Macclesfield, Cheshire, SK11 9DL, UK\\
 $^{2}$Dept. of Physics and Astronomy, University of British Columbia, 
 6224 Agricultural Road,Vancouver, BC V6T 1Z1, Canada}

\begin{abstract}
Some of the issues concerning polarisation calibration in the context
of pulsar observations are outlined along with the existing
calibration system in Jodrell Bank Observatory. The principle behind
the back-calibration of the Jodrell Bank database of 16 years is
presented along with plans for exploiting the database in the future.
\end{abstract}

\section{Introduction}

Polarisation can prove to be an invaluable tool in determining, among
other things, the geometrical properties of objects and phenomena. A large
amount of information can be retrieved from studying pulsar 
polarisation profiles. It can also be used to probe the emission
mechanism and geometry of pulsars, pulsar precession and
the magnetic field of our Galaxy. 

The Jodrell Bank database of polarisation observations
covers more than 16 years and over 300 pulsars observed at
frequencies of 230, 400, 600, 920, 1400 and 1600 MHz (Gould \& Lyne
1998), providing an ideal archive for thorough polarisation research.  
 
\section{Necessity of calibration: the case of Jodrell Bank}

Acquiring accurate polarisation profiles is not a trivial
task and a considerable number of calibration issues needs to be
addressed before the corrected profiles can be actually studied (e.g.
Hamaker, Bregman, \& Sault 1996 or Stinebring et al 1984). Some of the
main problems are: 

\begin{itemize}

\item Gain calibration

\item Feeds' non-orthogonality and ellipticity

\item Parallactic angle correction due to Earth's rotation

\item Faraday depolarisation due to the ionosphere \& the ISM

\item Signal path length difference

\end{itemize}

The calibration process at Jodrell Bank is based on a noise diode.
In particular, information from the 4 polarimeter channels with the
diode being alternatively switched on and off is used. This is combined with
observations of an unpolarised continuum source of known flux density
(e.g. 3C123) and observations of 1 degree off the pulsar and at half
power.  

The profiles resulting from this procedure are gain calibrated,
they have any channel offsets subtracted and they are
corrected for parallactic angle rotation as well as the noise diode
position angle. However, effects related to the feeds misalignment and
depolarisation from Faraday Rotation and path differences in the
signal may still produce intrumental errors. 

Therefore, although the calibration already performed is adequate for
timing studies, more calibration observations are required to correct
for the above effects as well. Such observations are not always available 
in the archive though and in order to minimise the gaps in the data, a
method has to be developed that will compensate for the data that have
no dedicated calibration measurements.

\section{Back Calibration Method}

Gould \& Lyne (1998) have created average profiles for all the pulsars
in the Jodrell Bank archive that have polarisation information. These
integrated profiles were created by averaging all the
observations of a pulsar for a certain frequency and they are assumed 
to be the ''true'' pulse profile at that frequency for the purposes of
the Back Calibration Method. Examples of such profiles can be
seen in Fig. 1.

\begin{figure}[t]
\plotfiddle{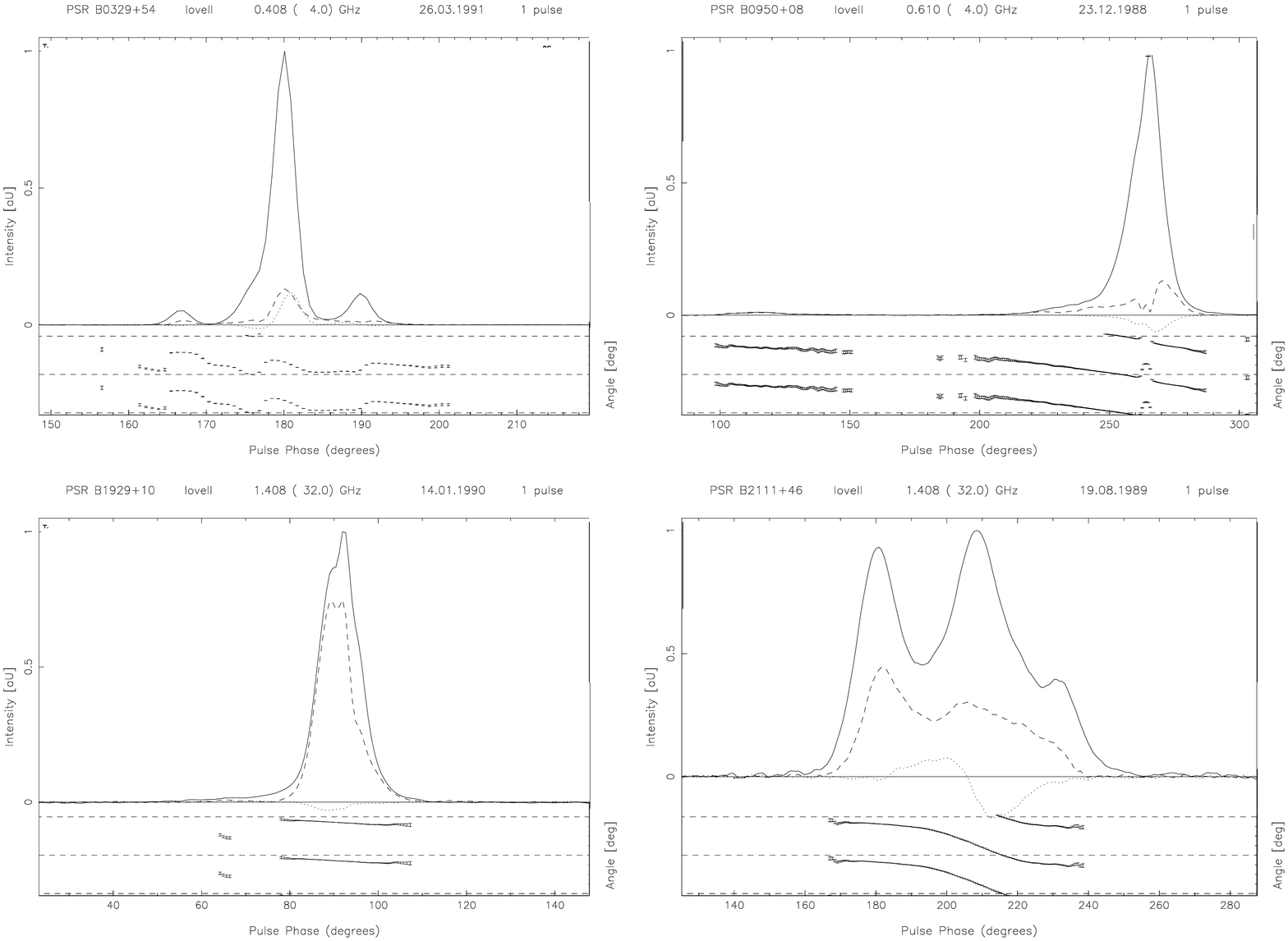}{0cm}{0}{44}{55}{-190}{-335}
\vspace{11.5cm}
\caption{A small example of the high signal to noise ratio of the
Gould \& Lyne archive of integrated profiles. Top Left: PSR B0329$+$54
at 408MHz, Top Right: PSR B0950$+$08 at 610 MHz, Bottom Left: PSR
B1929$+$10 at 1408 MHz, Bottom Right: PSR B2111$+$46 at 1408 MHz.}
\end{figure}

The main principle behind the Back Calibration Method is using
formalisation similar to that found in Johnston (2002). For a certain
time period when the telescope is observing using a certain 
system setup, the whole observing process introduces errors in the 
measurement such as those described in the previous section. To
determine them, the telescope, the polarimeter and whatever 
electronics intervene from the point the signal enters the system to
the point it is recorded are represented - to a first
approximation - by a 4$\times$4 matrix (the so-called Mueller Matrix - M). 
In reality, the 16 M elements can be reduced to 8 as they are interrelated.

Therefore, the following would hold:
\begin{equation}
      S^{\prime}=M\cdot S
\end{equation}
where $S^{\prime}$ is the Stokes Parameters' 4-vector which,
when combined for all bins, will make up an observation from the
same system setup period whose M is to be determined. S is the corresponding
corrected (or 'true') vector.  

Thus, the M fully describing a certain system setup can be determined 
from Eq. 1 and a certain observation, if the 'true' integrated
profiles by Gould \& Lyne (1998) are used. In particular, in order to calculate the M
elements, the observation and the true profile are compared and one
vector equation like Eq. 1 is formed for each bin that
has a signal to noise ratio of over 7. Further equations can be formed if one 
compares observations from different times of day (corresponding to
various parallactic angles) or from different pulsars (but always from
the same system setup) with the respective true profile found in the archive.

The set of equations thus formed can then be solved for the 8 unknown elements
(e.g. using a SIMPLEX method) determining in that manner M. The
more equations used in the fitting process the more accurate M
is likely to be. It then follows that:
\begin{equation}
      S=M^{-1}\cdot S^{\prime}
\end{equation}

Therefore, M can be used to retrieve the true Stokes Parameters
for that particular observation from Eq. 2. The product
of the inverse M times the observed Stokes Parameters will be a
profile with any instrumental effects removed. 

The process relies on the assumption that the archive profiles of
Gould \& Lyne are true representatives of the actual pulsar
profiles. However, even if some of the archive profiles used are not
accurate, incorporating as many as possible in the Back Calibration
Method and following a selection of only the higher quality ones,
ensures the required accuracy when used in conjunction with a
least-squares method.

\section{Future prospects}

As soon as a procedure is fully established with which any observation
existing in the database can be calibrated using information drawn
from the database itself, a number of projects can then start to be
examined.

For example, the precession of PSR B1828$-$11 can be verified and
studied by looking at the behaviour of its PA swing through the
15-year long baseline (Stairs et al these proceedings). In addition
to this, all 300 pulsars in the archive can be examined as to if they
exhibit similar precession characteristics. Even outside the scope of
precession, the study of the evolution of the PA swing through time
(especially in such long scales) is of considerable interest on its own.


As mentioned above, the geometry of the pulsar emission 
can also be studied. This is feasible by fitting for the PA swings and thus
determining the impact parameter $\beta$ (i.e. the angle between the
magnetic field axis and the observer's line of sight) and the angle
$\alpha$ (between the magnetic axis and the rotation axis).
Studying the emission geometry can provide constraints for emission
models and help answer questions regarding emission regions and
origin, single pulse structure and giant pulses.

Finally, it must be noted that the existing database will be extended
with the use of COBRA (Coherent Online Baseband Receiver for
Astronomy) the new Beowulf cluster of 182 processors. A more
systematic and careful calibration is implemented from the
beginning to COBRA which will allow for up to 100 MHz total B/W of
coherently dedispersed and fully calibrated data with the 4 Stokes
parameters of polarisation provided 'for free' - i.e. determined in
software). Such an instrument can be used in a wide number of
observations where high profile resolution is necessary or to address
entirely new questions such as single pulse polarimetry of Millisecond
Pulsars or MSP Rotation Measures. 

\acknowledgments DA would like to thank Bhal Chandra Joshi and
Christine Jordan for valuable discussions as well as Anita Richards
without whose help the poster would not have made it to the conference.
DA acknowledges the support of PPARC for the funding of this research
project. IHS is supported by an NSERC UFA and Discovery Grant.

\section{References}
Gould, D. M. \& Lyne, A. G. 1998 in \mnras, {\bf 301}, 1, pp. 235-260\\
Hamaker, J. P., Bregman, J. D., Sault, R. J. 1996 in \aaps, {\bf 117},
p.137-147\\
Johnston, S. 2002 in Publications of the Astronomical Society of
Australia, {\bf 19}, 2, pp. 277-281\\
Stairs, I. H., Lyne, A. G., Athanasiadis D., \& Kramer M. these proceedings\\
Stinebring, D. R., Cordes, J. M., Rankin, J. M., Weisberg, J. M., \&
Boriakoff, V. 1984 in \aaps, {\bf 55}, June 1984, p. 247-277\\
\end{document}